\documentclass[aps,prc,twocolumn,showkeys,showpacs]{revtex4}
\usepackage{amssymb}
\usepackage{amsmath}
\usepackage{graphicx}
\usepackage{color}

\newcommand{\req}[1]{(\ref{#1})}
\newcommand{\bel}[1]{\begin{equation}\label{#1}}
\newcommand{\belar}[1]{\begin{eqnarray}\label{#1}}
\def\rama{\vadjust{\vbox to 0pt{\vss \hbox to \hsize{\hskip\hsize
$\Leftarrow new $\hss}\vskip 3.5pt}}}
\def\mev{\;{\rm MeV}}
\def\pr{\prime}

\def\eps{\epsilon}
\def\epsk{\epsilon_k}
\def\del{\delta}

\begin{document}

\title{The Scission-Point Configuration within the Two-Center Shell Model Shape Parameterization}

\author{F.A.~Ivanyuk}
\email{ivanyuk@kinr.kiev.ua}
\affiliation{Tokyo Institute of Technology, Tokyo, Japan,\\ Institute for Nuclear Research, Kiev, Ukraine}
\author{S. Chiba}
\email{chiba.satoshi@nr.titech.ac.jp}
\affiliation{Tokyo Institute of Technology, Tokyo, Japan}
\author{Y.Aritomo,}
\email{aritomo.yoshihiro@nr.titech.ac.jp}
\affiliation{Tokyo Institute of Technology, Tokyo, Japan}

\date{\today}
\begin{abstract}
Within the two-center shell model parameterization we have
defined the optimal shape which fissioning nuclei attain just
before the scission and calculated the total deformation energy
(liquid drop part plus the shell correction) as function of the
mass asymmetry and elongation at the scission point.
The three minima corresponding to mass symmetric and two mass
asymmetric peaks in the mass distribution of fission fragments
are found in the deformation energy at the scission point. The
calculated deformation energy is used in quasi-static
approximation for the estimation of the total kinetic and excitation energy
of fission fragments and the total number of emitted prompt neutrons. The
calculated results reproduce rather well the experimental data on
the position of the peaks in the mass distribution of fission
fragments, the total kinetic and excitation energy of fission
fragments. The calculated value of neutron multiplicity is
somewhat larger than experimental results.
\end{abstract}


\maketitle

\section{Introduction}
\label{intro}
In the theory of nuclear fission the quasistatic quantities
like the potential energy surface, the ground sta\-te energy and deformation, the fission barrier height, etc.,
are often calculated within the macroscopic-microscopic method \cite{Str66,brdapa}. In this method the total energy of the fissioning nucleus consists of the two parts, macroscopic and microscopic. Both parts are calculated at fixed shape of nuclear surface. In the past a lot of shape parameterizations were proposed and used.
A good choice of the shape parameterization is often the key to the
success of the theory. Usually, one relies on physical intuition
for the choice of the shape parameterization.

A method to define the shape of nuclear surface which does
not rely on any shape parameterization was proposed by V.
Strutinsky in \cite{stlapo,strjetp45}. In this approach the shape
of an axial, left-right symmetric nucleus was defined by looking
for the minimum of the liquid-drop energy under the additional
restrictions that fix the volume and elongation of the  drop.

Recently the method was further developed \cite{fivan08,ivapom2} by incorporating of the axial \cite{ivpoba,ivapom4} and left-right asymmetry and the neck degree of freedom of the nuclear shape \cite{procedia,fivan14}.

The important result of the Strutinsky procedure \cite{stlapo} is the possibility to definite in a formal way the scission point as the maximal elongation at which the nucleus splits into two fragments.

Having at one's disposal the shape and the deformation energy at
the scission point in present work we have tried to evaluate the measurables of the
fission experiments like mass distribution, the total kinetic and
excitation energy of fission fragments, the multiplicity of
prompt neutrons.

A similar investigation was carried out in a recent work
\cite{cahaivta} where the experimental results for
$^{235}{\text U}+n_{th}$ reaction were described in terms of three
fission modes.

The paper is organized as follows. Section
\ref{optimal} contains a short overview of Strutinsky
prescription \cite{stlapo} for the optimal shapes. The shapes
that correspond to the minimum of liquid-drop energy within the
two-center shell model shape parameterization are presented in
Section \ref{opttcsm}. The account of the shell correction and
their influence on the total excitation energy of fission
fragments are discussed in Section \ref{shells}. The comparison of
the calculated and experimental results for the total kinetic energy of fission fragments and multiplicity
of prompt neutrons for the fission of $^{232}{\rm Th}, ^{235}{\text U},
^{239}{\rm Pu}$ and $^{245}{\rm Cm}$ by thermal neutrons is given in Section \ref{observa}. Sect. \ref{summa} contains short summary.

\section{Optimal shapes}
\label{optimal}
In \cite{stlapo}  the shape of a
left-right and axial symmetric nucleus was described by some profile function
$\rho(z)$. The shape of the surface was obtained then by rotating
the $\rho(z)$ curve around the $z$-axis. A formal definition
of $\rho(z)$ was obtained by searching for the minimum of the
liquid-drop energy, $E_{\rm LD}=E_{\rm surf} + E_{\rm Coul}$,
under the constraint that the volume $V$ and the elongation
$R_{12}$ are fixed,
 \bel{variation}
 \frac{\delta}{\delta \rho}\left[E_{\rm LD} - \lambda_1 V -\lambda_2 R_{12} \frac{}{}\! \right] = 0 ,
\end{equation}
with
 \bel{r12}
 V=\pi\int\limits_{z_1}^{z_2}\rho^2(z) dz \,\,,
\quad  R_{12}=\frac{2\pi}{V}\int\limits_{z_1}^{z_2}\rho^2(z)\vert z\vert dz\,\,.
\end{equation}
In \req{variation} $\lambda_1$ and $\lambda_2$ are the
corresponding Lagrange multipliers. The elongation parameter
$R_{12}$ was chosen in \cite{stlapo} as the distance between
the centers of mass of the left and right parts of the nucleus, see \req{r12}. 

Since both Coulomb
and surface energy
are functionals of $\rho(z)$ the variation in \req{variation} results in an
integro-differential equation for $\rho(z)$
 \bel{diffeq}
\rho\rho^{\prime\prime} \!= 1+(\rho^{\prime})^2 - \rho [ \lambda_1 +
\lambda_2 \vert z
 \vert
       - 10x_{\rm LD}\Phi_S ] \! \left[1+(\rho^{\prime})^2 \right]^\frac{3}{2}.
\end{equation}
Here $\Phi_S\equiv\Phi(z, \rho(z))$ is the Coulomb potential at
the nuclear surface and $x^{}_{\rm LD}$ is the fissility parameter
of the liquid drop \cite{bohrwhee},
\begin{equation}\label{xldm}
x_{\rm LD}\equiv \frac{E_{\rm Coul}^{(0)}}{2E_{\rm surf}^{(0)}}
   =\frac{3}{10}\frac{Z^2e^2}{4\pi R_0^3\sigma}\approx \frac{Z^2}{49A}\,\,,
\end{equation}
where $\sigma$ is the surface tension coefficient, see Eq.\ (\ref{es0ec0}).
In (\ref{xldm}) and everywhere below the index $^{(0)}$ refers to the spherical shape.

\begin{figure}[ht]
\includegraphics[width=0.9\columnwidth]{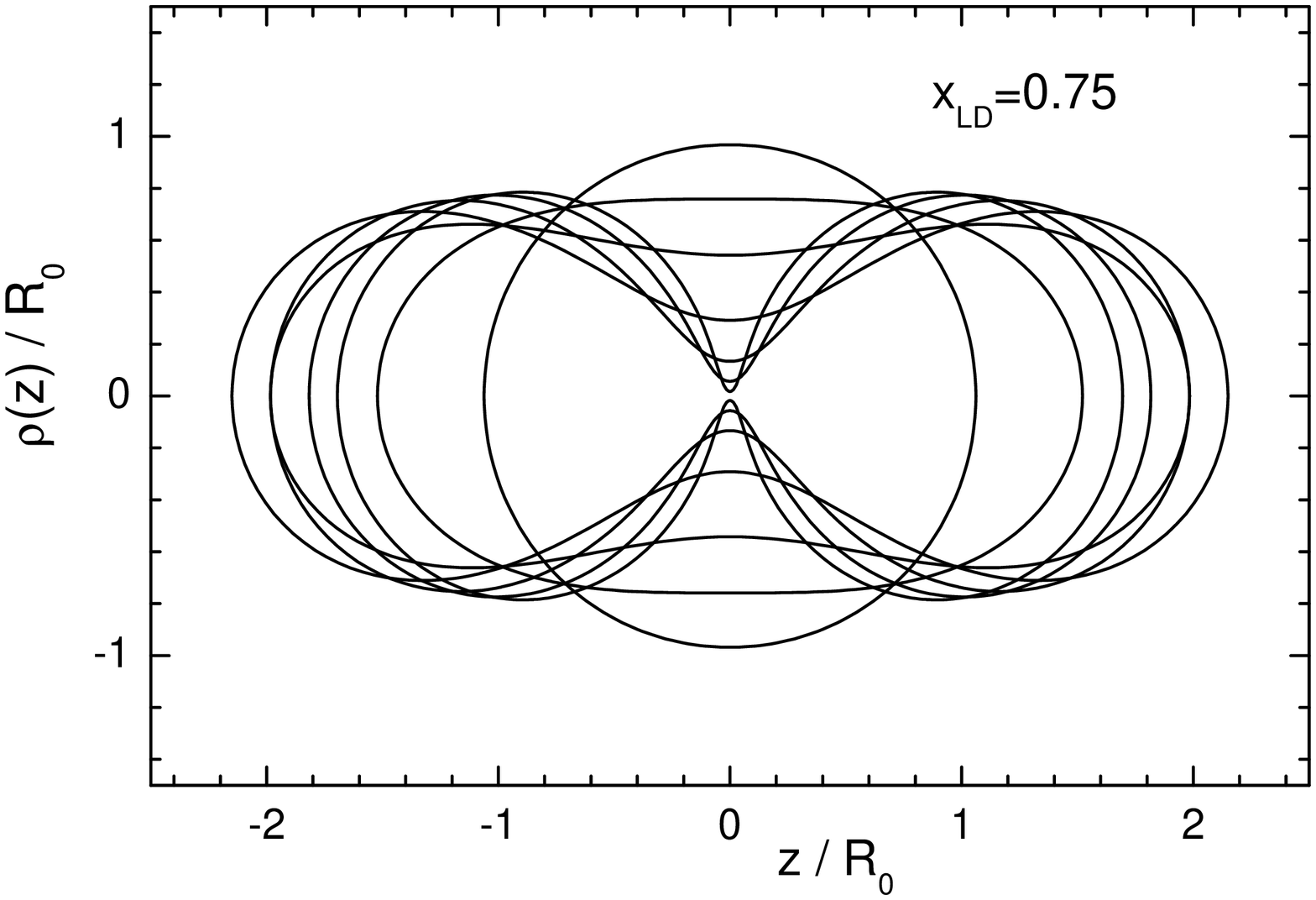}
\caption{\label{os} Solutions of Eq. \protect\req{diffeq} for a few values of Lagrange multiplier $\lambda_2$.}
\end{figure}

By solving Eq. (\ref{diffeq}) for given $x_{LD}$ and $\lambda_2$
($\lambda_1$ is fixed by the volume conservation condition) one
obtains the profile function $\rho(z)$ which we refer to as the
{\it optimal shape}. Varying parameter $\lambda_2$   one obtains
a full variety of shapes ranging from a very oblate shape (disk,
even with central depression) up to two touching spheres. Few
examples of optimal shapes are shown in Fig. \ref{os}.

\begin{figure}[ht]
\includegraphics[width=\columnwidth]{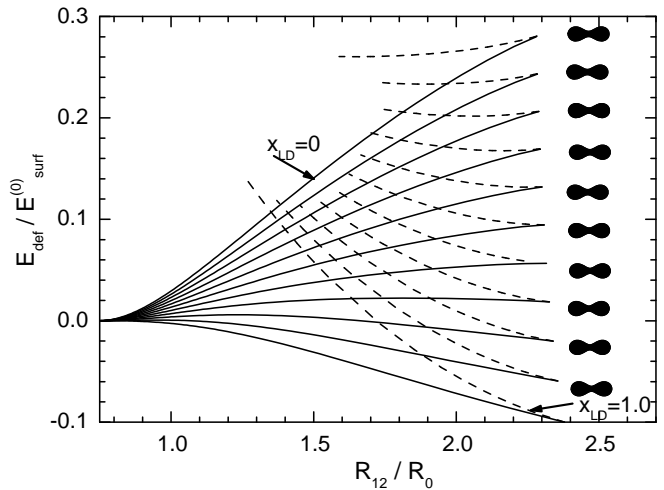}
\caption{\label{edef} Liquid-drop deformation energy
(\protect\ref{eldm}) as function of parameter $R_{12}$ for a few
values of the fissility parameter $x_{\rm LD}$.}
\end{figure}

The liquid drop deformation energy $E_{\rm def}^{LD}=E_{\rm LD}-E_{\rm LD}^{(0)}$
\bel{eldm}
E_{\rm def}\equiv E_{\rm def}^{LD}/{E_{\rm surf}^{(0)}}=B_{\rm surf}-1 + 2x_{\rm
LD}(B_{\rm Coul}-1)\,,
\end{equation}
(in units  of the surface energy for a spherical shape)
\begin{equation}\label{es0ec0}
E_{\rm surf}^{(0)}=4\pi\sigma R_0^2,\quad
E_{\rm Coul}^{(0)}={3}{Z^2e^2}/{5R_0}\,\,.
\end{equation}
calculated with optimal shapes is shown in Fig.~\ref{edef} as function of the elongation parameter $R_{12}$. Note, that the spherical shape corresponds to $R_{12}=0.75$.
In \req{eldm}
\bel{bcbs}
B_{\rm Coul}\equiv {E_{\rm
Coul}}/{E_{\rm Coul}^{(0)}}\quad {\rm and}\quad
B_{\rm surf}\equiv {E_{\rm surf}}/{E_{\rm surf}^{(0)}} ,
\end{equation}
where $R_0$ is the radius of the spherical nucleus.

One can see from Fig.~\ref{edef} that the elongation $R_{12}$ of
the shapes shown in these figures is limited by some maximal
value $R_{12}^{(crit)}$. With a good accuracy
the maximal deformation is independent of the fissility parameter
$x_{\rm LD}$, $2.32\leq R_{12}^{\rm max}/R_0\leq 2.35$ for $0.4\leq
x_{\rm LD}\leq 0.9$. Above this deformation mono-nuclear
shapes do not exist.  This maximal deformation was interpreted in
\cite{stlapo} as the scission point.

The different branches of the energy shown in Fig. \ref{edef}
correspond to the different values of fissility parameter
$x_{LD}$. Along each lower branch the shape of the drop changes
from sphere to spheroid, then to the shape with the neck which is
getting smaller until the maximal elongation $R_{12}^{\rm max}$,
see Fig.~\ref{os}. The shapes at the maximal elongation
are shown in the right part of Fig.~\ref{edef}. It turns out that
at each fixed $R_{12}$ the shape itself does not depend much on
the fissility parameter.

Another peculiarity of Fig.~\ref{edef} is the upper branches of the deformation
energy at large deformation. Along these branches
the neck of the drop becomes smaller and smaller until the shape turns into two
touching spheres.

It turns out, however, that the upper branch of the deformation energy
corresponds not to the {\it minimum} of $E_{\rm LD}$ but to its {\it maximum} (Eq.\req{variation} holds true both at the minimum and maximum of the liquid-drop energy).
This can be easily verified by adding to the optimal profile function $\rho(z)$ some
small perturbation $\delta \rho(z)$ and calculating the deformation energy with
perturbed profile function $\rho(z)+\delta \rho(z)$. Thus, the upper branch of $E_{\rm
def}$ corresponds to the ridge of the potential energy surface in the coordinates
elongation and neck thickness.

\section{The optimal shapes within two-center shell model parameterization}
\label{opttcsm}
The two-center shell model parameterization (TCSMP)
\cite{maruhn} is at present the only parameterizations which describes {\it
both} compact shapes and separated fragments. It is successfully used
 by few groups in the description of fission-fusion
process based on the Langevin equations for the shape parameters,
see \cite{wada,aritomo,arichi,antonenko} and references therein.

\begin{figure}[ht]
\includegraphics[width=0.8\columnwidth]{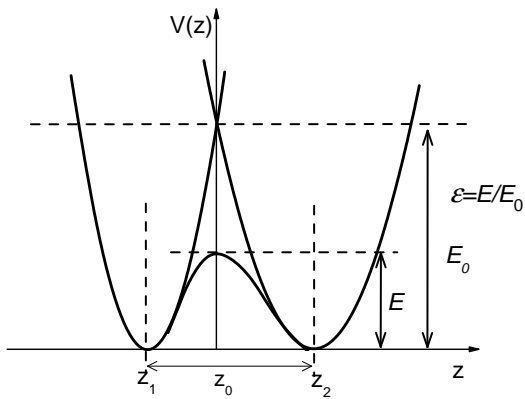}
\caption{\label{tcsmp} The mean field potential of the two-center shell model.}
\end{figure}

The two-center shell model potential includes the central part $V(\rho, z)$, ${\bf ls}$ and ${\bf l}^2$ terms. The central part consists of two oscillator potentials smoothly joined by the fourth order polynomial, see Fig. \ref{tcsmp}. 
The sharp surface $\rho(z)$ is defined as that
given by the equipotential surfaces of potential $V(\rho, z)$,
i.e. by the equation $V(\rho(z), z)=V_0$.
The constant $V_0$ is
found from the requirement that the volume inside sharp
surface is equal to the volume of spherical nucleus.

Within TCSMP the
shape is characterized by 5 deformation parameters: the distance
$z_0$ between the centers of left and right oscillator potentials,
the mass asymmetry parameter $\alpha\equiv (A_H-A_L)/(A_H+A_L)$, the deformations $\del_1$
and $\del_2$ of the left and right oscillator potentials and the neck
parameter $\eps$.
Here $A_H$ and $A_L$ are the masses of heavy and light fragments (in case of the shape separated into two fragments) or the masses of the right and left parts of the compact nucleus. Within TCSMP the shape is divided in parts by the point $z=0$. Please, note, that $\delta_1$ and $\delta_2$ fix the deformation of potential only in "outer" region, namely for $z\leq z_1$ or $z_2\leq z$. The deformation of whole fragments depends not only on $\delta_1$ and $\delta_2$ but on all other parameters, $z_0, \epsilon$ and $\alpha$.

By solving the Langevin equations the number of parameters is
often reduced to 3 in order to diminish the computation time. The
parameters $\del_1$ and $\del_2$ are assumed to be the same
$\del_1=\del_2$, and the neck parameter $\eps$ is kept constant.
Here we would like to note, that constant $\eps$ does not mean
constant neck radius. Within TCSMP the neck radius depends on all
5 deformation parameters. Even by variation of only one
deformation parameter $z_0$ one gets quite reasonably class of
shapes of fissioning nucleus.

\begin{figure}[ht]
\includegraphics[width=\columnwidth]{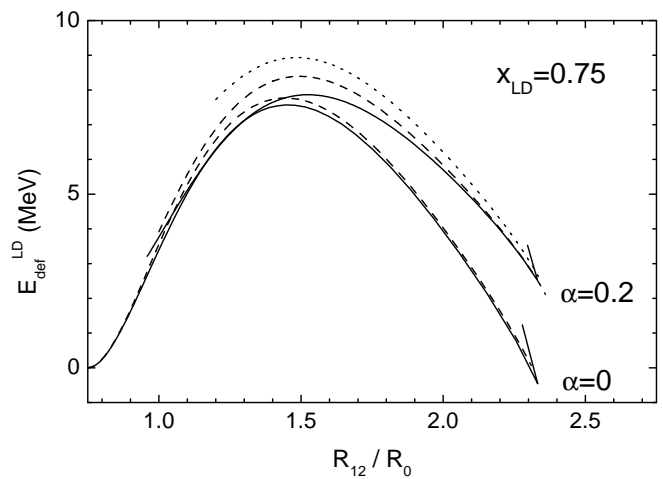}
\caption{The liquid-drop energy \protect\req{edef} calculated
with the optimal shapes \protect\req{diffeq} (solid line) and
with the two-center shell model parameterization (dash line)
minimized with respect to $z_0, \eps$ and the fragments
deformation $\delta_1, \delta_2$ for the mass asymmetry
$\alpha=0$ and $\alpha=0.2$ as function of the of the distance
$R_{12}$ \protect\req{r12} between centers of mass of left and
right parts of the drop for the fissility parameter
$x_{LD}=0.75$. The dot line shows the result of minimization
within TCSM parameterization with $\delta_1=\delta_2$
restriction} \label{accura}
\end{figure}

Having at one's disposal the optimal shapes is
of a certain interest to check how good is the two-center shell
model parameterization. For this purpose we compare in
Fig.~\ref{accura}  the liquid-drop deformation energy
calculated with the optimal shapes \req{diffeq} and the
two-center shell model parameterization for the symmetric splitting ($\alpha=0$) and mass asymmetry $\alpha=0.2$ that corresponds to the position of the main peak in the mass distribution of fission fragments of considered in this work nuclei .
The deformation energy is plotted as a function of $R_{12}$
\req{r12}. In case of two-center shell model parameterization the
deformation energy was minimized with respect to $z_0, \eps$, $\del_1$ and
$\del_2$ keeping constant the distance $R_{12}$ between the centers of mass of left and right part of nucleus.

As one can see from Fig.~\ref{accura}, the TCSMP is rather
accurate. In case of symmetric splitting  ($\alpha=0$) the
difference between the deformation energy calculated with optimal
shape and TCSMP does not exceed 0.4 $\mev$. The optimal value of
parameter $\eps$ varies somewhat with $z_0$ and at large
deformations is close to $\eps\approx 0.25$. For the mass
asymmetry $\alpha=0.2$ the difference between the deformation
energy calculated with optimal shape and TCSMP is somewhat
larger. In the barrier region it is of the order of 0.8 $\mev$.

At present and in the past a lot of calculations were done with the restricted set of TCSM shapes, namely putting $\delta_1=\delta_2$. In order to check how good is this approximation we have carried out the minimization of the liquid-drop energy within TCSMP with $\delta_1=\delta_2$ restriction (dot line in Fig. \ref{accura}). In this case the difference between the deformation energy calculated with optimal shape and TCSMP is of the order of 1.6 $\mev$ at the barrier.

At the scission point the difference between the deformation
energy calculated with optimal shape and TCSMP is very small, it
does not exceed few hundred $keV$ even in $\delta_1=\delta_2$
case. Since in present work we examine the deformation energy at
the scission point configuration the use of TCSMP seems quite
justified.

\begin{figure}[ht]
\includegraphics[width=\columnwidth]{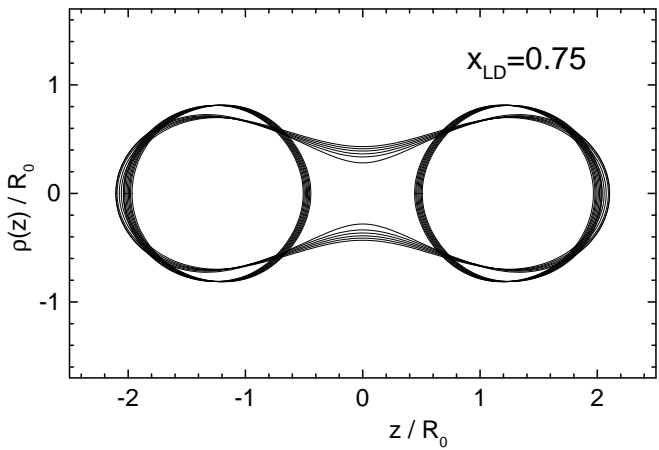}
\caption{The change of the shape of the drop around the scission point.}
\label{scission}
\end{figure}

Since the TSCMP describes both compact and separated shapes, it is
interesting to check what will be the optimal shape within TSCMP
beyond $R_{12}^{(crit)}$. For this purpose  we carry out
the minimization of the liquid drop energy within TCSMP with respect to $z_0,
\eps$ and $\del_1$ keeping $R_{12}$ fixed for $R_{12}>
R_{12}^{(crit)}$. The result is shown in Fig~\ref{scission}.
Instead of continuous decreasing of the neck to zero, the shapes
suddenly splits into two fragments at relatively thick neck,
$R_{neck}\approx 0.3 R_0$.

\begin{figure}[ht]
\includegraphics[width=0.8\columnwidth]{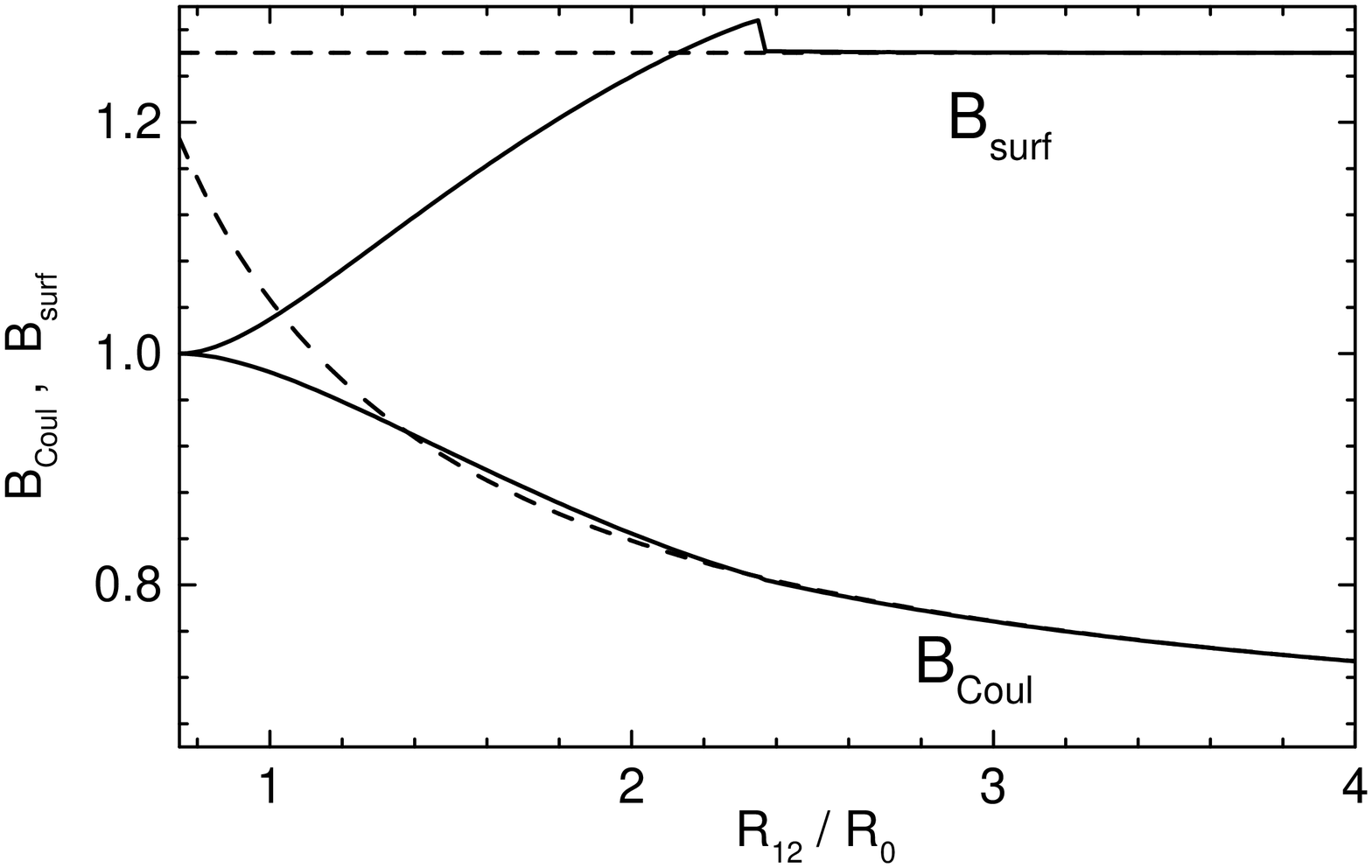}
\caption{The relative Coulomb and surface energies \protect\req{bcbs} as function of the distance $R_{12}$ \protect\req{r12} between centers of mass of left and right parts of the system calculated for the shapes shown in Fig. \protect\ref{scission} (solid) and for the two separated spheres (dash).}
\label{bsurf}
\end{figure}

The shape of the fragments "immediately after scission" is very
close to the spheres. This is confirmed by the comparison in
Fig.~\ref{bsurf} of the Coulomb and surface energy of left and
right parts of the drop with these of the two separated spheres.
Notice, that at the scission point the Coulomb energy changes
continuously, only the surface energy changes abruptly to the
smaller value.

\begin{figure}[ht]
\includegraphics[width=0.8\columnwidth]{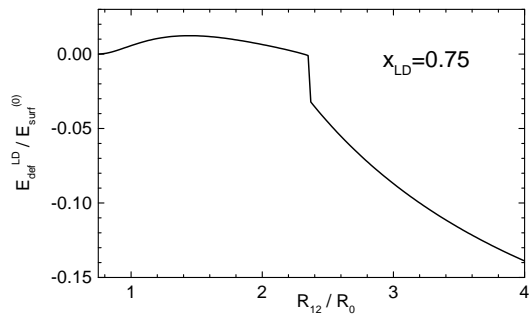}
\caption{The dependence of the liquid-drop energy \protect\req{eldm} on the distance $R_{12}$ between centers of mass of left and right parts of nucleus.}
\label{jump}
\end{figure}

The "jump" of the surface energy to smaller value at the scission point causes the corresponding jump in the liquid-drop energy, see. Fig.~\ref{jump}.

The deformation energy shown in Fig.~\ref{jump}, differs from the
published so far results by the presence of jump at critical
elongation. This jump is a consequence of using a special (though
very clear) quantity for measure of elongation of nucleus - the
distance $R_{12}$ between centers of mass of left and right parts
and requirement that
other deformation parameters are found from
the minimization of liquid-drop energy at given $R_{12}$. These
other parameters (the distance $z_0$, the neck parameter $\eps$
and the deformation of fragments) change abruptly at
$R_{12}=R_{12}^{(crit)}$ to ensure that the energy is minimal on
both side of $R_{12}^{(crit)}$. Without minimization one usually consider
the continuous change of the deformation parameters on the whole
potential energy surface.

The possibility to define  the critical deformation
$R_{12}^{(crit)}$ in a formal way has at least two important
consequences: it clearly indicates at which deformation the
dynamical calculations should be stopped, and it makes to
calculate the primary excitation energy of the fission fragments
during the scission. The difference of the deformation energy
"just before scission" ($jbs$) and "immediately after scission"
($ias$) \bel{deledef} \Delta
E_{def}^{LD}=E_{LD}^{(jbs)}-E_{LD}^{(ias)}
\end{equation}
can characterize the excitation energy available for emission of prompt neutrons and $\gamma$-quanta by the fission fragments.
Here by "just before scission" and "immediately after scission" we call the configuration with $R_{12}$ infinitely smaller or larger than  $R_{12}^{(crit)}$.
\section{The account of shell effects}
\label{shells}
For accurate calculation of the excitation energy released
in result of neck rupture the account of shell effects is very
important. Besides, one should consider the possible mass
asymmetry $\alpha$ of the fission fragments.
\begin{figure}[h]
\includegraphics[width=\columnwidth]{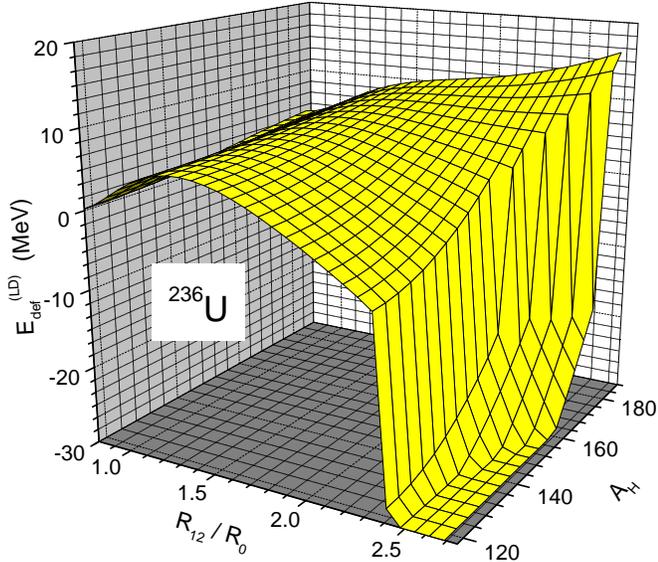}
\caption{(Color online) The liquid-drop deformation energy of $^{236}{\text U}$ as
function of the elongation $R_{12}$ \protect\req{r12} and the
mass number $A_H$ of heavy fragment.} \label{eldm236}
\end{figure}

\begin{figure}[ht]
\includegraphics[width=\columnwidth]{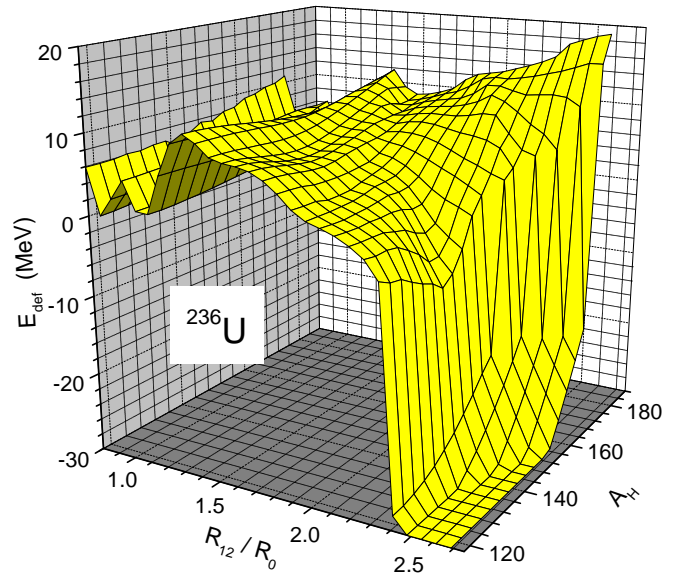}
\caption{(Color online) The total (liquid-drop plus shell correction)
deformation energy of $^{236}{\text U}$ as function of the elongation
$R_{12}$ \protect\req{r12} and the mass number $A_H$ of heavy
fragment.} \label{etot236}
\end{figure}

In Figs.~\ref{eldm236}-\ref{etot236} we show the liquid-drop and the total (liquid-prop plus the shell correction including the shell correction to the pairing energy)
\bel{etot}
E_{def}=E_{def}^{LD}+\delta E, \,{\rm with}\,\,\,
\delta E= \sum_{n,p}(\delta E_{shell}^{(n,p)}+\delta E_{pair}^{(n,p)})
\end{equation}
deformation energy for nucleus $^{236}{\text U}$. The summation in \req{etot} is carried out over the protons ($p$) and neutrons($n$).

The $\delta E_{shell}$ was defined in a usual way as the difference between the sum of single particle energies of occupied states $E_{ipm}$ and the averaged quantity,
\bel{shco}
E_{ipm}=\sum_{occ.}\epsk,\quad \delta E_{shell}=E_{ipm}-\int_{-\infty}^{\widetilde\mu}\,e\,\widetilde g(e)\,de .
\end{equation}
Here $\widetilde g(e)$ is the averaged density of single-particle states,
\bel{gavr}
\widetilde g(e)=\frac{1}{\gamma}
\int_{-\infty}^{\infty}f\left(\frac{e-e^{\pr}}{\gamma}\right)g_s(e^{\pr})de^{\pr}=\frac{1}{\gamma}\sum_k\,f\left(\frac{e-\epsk}{\gamma}\right),
\end{equation}
with $g_s(e)$ being the density of single-particle states,
\bel{gs} g_s(e)=\sum_k \del(e-\epsk).
\end{equation}
The chemical potential $\widetilde\mu$ is defined by the particle number conservation condition
\bel{mu}
\quad \int_{-\infty}^{\widetilde\mu}\,\widetilde g(e)\,de= N,
\end{equation}
where $N$ is the particle number (neutrons or protons).
The $f(x)$ is Strutinsky smoothing function \cite{brdapa},
\bel{fx}
f(x)=\frac{e^{-x^2}}{\sqrt{\pi}}\sum_{n=0,2,...}^{M}\alpha_nH_n(x),\,\alpha_0=1,\,\alpha_{n+2}=\frac{-\alpha_n}{2(n+2)}.
\end{equation}
In \req{fx} $H_n(x)$ are the Hermite polynomials. The smoothing
width $\gamma$ and the degree $M$ of correcting polynomial are
the parameters of the averaging procedure. They were fixed by the
so called "plateau condition".

The shell correction to the pairing energy was defined as the difference between the pairing energy $E_{pair}$ in BCS approximation,  $E_{pair}=E_{BCS}-E_{ipm}$,
\bel{ebcs}
E_{pair}=\sum_{k=k_1}^{k_2} (2v_k^2-n_k)(\epsk-\lambda)-\frac{\Delta^2}{G},
\end{equation}
and the averaged part $\widetilde E_{pair}$,
\bel{epair}
\delta E_{pair}=E_{pair}-\widetilde E_{pair}.
\end{equation}
The $\lambda$ and $\Delta$ in \req{ebcs} are fixed by the particle number conservation and the gap equation.

We define $\widetilde E_{pair}$ following \cite{brdapa}, i.e.  by replacing the sum in \req{ebcs} by the integral and assuming that the density of state is constant over the pairing window
\belar{epairavr}
\widetilde E_{pair}&=&\widetilde g(\lambda)\int_{\lambda-\hbar\Omega}^{\lambda+\hbar\Omega}(e-\lambda)  \left(1-\frac{e-\lambda}{\sqrt{(e-\lambda)^2+\widetilde\Delta^2}}\right)de\nonumber\\
&-&\frac{\widetilde\Delta^2}{G}-2\widetilde g(\lambda)\int_{\lambda-\hbar\Omega}^{0}(e-\lambda)d(e-\lambda)\\&=&\widetilde g(\lambda)(\hbar\Omega)^2[1-\sqrt{1+\widetilde\Delta^2/(\hbar\Omega)^2}]\approx -\frac{1}{2}\widetilde g(\lambda)\widetilde\Delta^2\nonumber.
\end{eqnarray}
The pairing strength $G$ was removed from \req{epairavr} by solving the gap equation in the same approximation,
\bel{gapavr}
\frac{1}{G}=-\widetilde g(\lambda){\rm ln}{[\sqrt{1+(\hbar\Omega)^2/\widetilde\Delta^2}-\hbar\Omega/\widetilde\Delta)]}.
\end{equation}
For the average pairing gap $\widetilde\Delta$ we used the approximation suggested in \cite{moller92}
\bel{ugap}
\widetilde\Delta=
\begin{cases}
         r\,e^{-tI^2+sI}/Z^{1/3}, \hspace{0.5cm} \text{for protons}, \\
         r\,e^{-tI^2-sI} /N^{1/3}, \hspace{0.5cm} \text{for neutrons},
\end{cases}
\end{equation}
with $r=5.72 MeV, s=0.118, t=8.12, I\equiv (N-Z)/A$.

The parameter $z_0$ of TCSMP can be expressed as a function of
$R_{12}, \alpha, \delta_1, \delta_2, \eps$. 
The 
parameters, $\eps, \delta_1, \delta_2$, were found by the
minimization of the liquid-drop energy at fixed $R_{12}$ and
$\alpha$.

For the calculation of the shell correction the optimal TCSMP shape was expanded in series in deformed Cassini ovaloids (up to 20 deformation parameters were included). For the shape given in terms of Cassini ovaloids the shell model code \cite{pash71} with deformed Woods-Saxon potential was used to calculate the single-particle energies and the shell correction $\del E$.

Please, note that the shape in the above calculations was defined
by the minimization of the {\it liquid-drop energy}. The shell
correction was added afterwards to the optimal liquid-drop
energy. Thus, the available class of shapes was defined by the
liquid-drop properties of system. The influence of shell
structure on the available class of shapes was ignored. For
example, the configuration with one fragment being almost
spherical and another very elongated due to the shell effects is
not possible to obtain by the minimization of liquid-drop energy.
Such configuration leads to a second mass-asymmetric minimum in
the potential energy surface of actinide nuclei and has a
considerable influence on the mass distribution of fission
fragments.

In principle, one could minimize the total
(liquid-drop plus shell correction) energy within TCSM
parameterization keeping fixed $R_{12}$ and $\alpha$. However,
such procedure is very time consuming.
 That is why,
below we follow more closely the optimal shapes procedure
\cite{fivan14} and introduce an additional freedom for the shape of fissioning nucleus by minimization of the liquid-drop energy with additional constraints
\bel{econstr} E=E_{LD}-\lambda_1 V-\lambda_2 {\widetilde R}_{12}-\lambda_3 {\widetilde\delta} -\lambda_5 Q_{2L}-\lambda_6 Q_{2R}.
\end{equation}
Here $Q_{2L}$ and $Q_{2R}$ are the quadrupole moments of the left
and right parts of nucleus. The ${\widetilde R}_{12}$ and ${\widetilde\delta}$ are the  smoothed constraining operators for the elongation and mass asymmetry \cite{fivan14}.

The energy
\req{econstr} was minimized numerically with respect to
parameters $\delta_1, \delta_2$ and $\eps$ keeping fixed the
elongation $R_{12}$, mass asymmetry $\alpha$ (or mass of heavy
fragment $A_H$) and the Lagrange multipliers $\lambda_5,
\lambda_6$. Thus, the maximal elongation
$R_{12}=R_{12}^{(crit)}$, the shape and the energy at
$R_{12}^{(crit)}$ become dependent on $\lambda_5$ and
$\lambda_6$.

\begin{figure}[ht]
\includegraphics[width=\columnwidth]{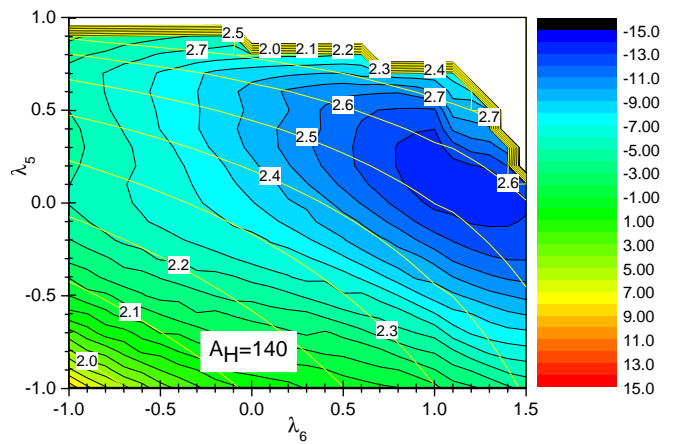}
\caption{(Color online) The total deformation energy at the
maximal elongation  $R_{12}=R_{12}^{(crit)}$ for the asymmetry
corresponding to heavy fragment mass $A_H=140$ calculated with
the solutions of Eq.\protect\req{econstr} as function of Lagrange
multipliers $\lambda_5$ and $ \lambda_6$, see
Eq.\protect\req{econstr}. The yellow contour lines show the value
of $R_{12}^{(crit)}$.} \label{lam56}
\end{figure}

In Fig. \ref{lam56} we show the maximal elongation
$R_{12}^{(crit)}$ and the {\it total} deformation energy at
$R_{12}=R_{12}^{(crit)}$ for $A_H=140$ as function of $\lambda_5$
and $\lambda_6$.

First of all, one can see that due to the shell effects the scission point configuration with the lowest total energy corresponds not to $\lambda_5=\lambda_6=0$ but to some finite values of $\lambda_5$ and $\lambda_6$. 

In order to show the dependence of the total energy (and the
shell correction $\delta E$) on the maximal elongation
$R_{12}^{(crit)}$, the lowest energy at fixed $R_{12}^{(crit)}$
(the point with the lowest energy along yellow contour line in
Fig. \ref{lam56}) from the total energy surface shown in Fig.
\ref{lam56} was picked  up. After such minimization we get the
total energy and the shell correction $\delta E$ as function of one variable
$R_{12}^{(crit)}$. This procedure was carried out for each value
of the mass asymmetry. In this way, the optimal scission point in $\{\lambda_5,
\lambda_6 \}$ space was found by the  minimization of the {\it
total} energy $E_{LD}+\delta E$ on the class of shapes given by the solution of Eq.
\req{econstr}.



The additional constraints  $\lambda_5 Q_{2L}$ and $\lambda_6
Q_{2R}$ were introduced in order to find the scission shape that
corresponds to the smallest total (liquid-drop plus shell
correction) energy. Thus, the configuration space in $\lambda_5,
\lambda_6$ should be large enough to include the point of minimal
total energy. For each value of mass asymmetry we have checked if
the point of minimal total energy is inside of chosen $\lambda_5,
\lambda_6$\, - space. If such point turned out to be at the edge of
$\lambda_5, \lambda_6$\, - space, the limiting values of $\lambda_5$
or $\lambda_6$ were increased.  Once the point of minimal total
energy is inside of chosen $\lambda_5, \lambda_6$\, - space, the
results of calculation are not sensitive to the size of
$\lambda_5, \lambda_6$\  - space since the contributions of the points
away from minimum are suppressed by the exponential factor
in \req{boltz}.

\begin{figure}[ht]
\includegraphics[width=\columnwidth]{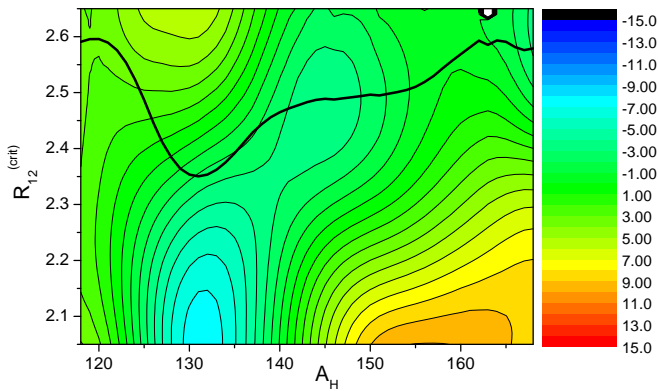}
\caption{(Color online) The shell component of the scission point deformation
energy of $^{236}{\text U}$ as function of the heavy fragment mass number
$A_H$ and the maximal elongation $R_{12}^{(crit)}$.  The mean
value \protect\req{rmean} of  $R_{12}^{(crit)}$ is shown by thick line.} \label{eshell}
\end{figure}

\begin{figure}[ht]
\includegraphics[width=\columnwidth]{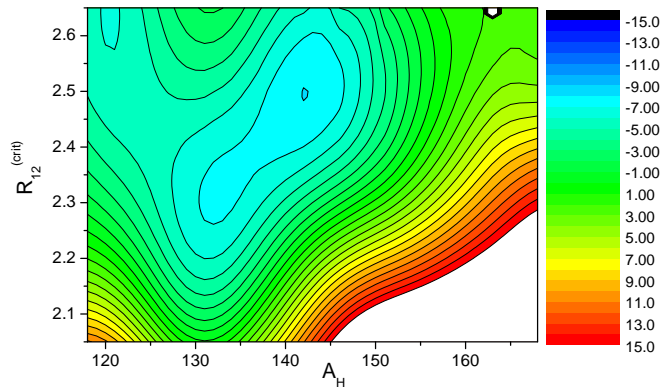}
\caption{(Color online) Total energy (liquid drop plus shell correction) for $^{236}{\text U}$
at the scission point as function of the heavy fragment mass
number $A_H$ and the maximal elongation $R_{12}^{(crit)}$.} \label{etot236b}
\end{figure}

The shell correction and the total deformation energy of
$^{236}{\text U}$ are shown in
Figs. \ref{eshell}, \ref{etot236b} as function of the heavy
fragment mass number $A_H$ and the elongation $R_{12}^{(crit)}$ at critical deformation.
The three minima, two at mass asymmetric deformation  $A_H\approx
141$ and $A_H\approx 134$ and one at symmetric deformation are
clearly seen in the deformation energy shown in Figs.
\ref{eshell}, \ref{etot236b}. The minimum at $A_H\approx 141$ is
responsible for the main peak in the mass distribution of the
fission fragments in the reaction $^{235}{\text U}+n_{th}$. The second
mass asymmetric minimum contributes to the satellite in the mass
distribution.
This minimum gives the main contribution to
the maximum of the total kinetic energy (TKE) of fission fragments at
$A_H\approx 130$. The reason is clear looking at mean value
\bel{rmean} \langle
R_{12}^{(crit)}(\alpha)\rangle=\sum_i R_{12}^{(crit)}(\alpha,
q_i)P(\alpha, q_i)/\sum_i P(\alpha, q_i)\,
\end{equation}
shown in Fig. \ref{eshell} by heavy solid curve. The
$<R_{12}^{(crit)}>$ is the most probable distance between the
centers of mass of left and right parts of nucleus at critical
deformation (just before scission). The averaging in \req{rmean}
is done with the canonical distribution in the space of
variables $\lambda_5, \lambda_6$, see \req{boltz} below.

The Coulomb repulsion energy
and, consequently, the total kinetic energy of fission fragments
is defined mainly by the distance between the centers of mass of
fragments. At the asymmetries where $<R_{12}^{(crit)}>$ is
maximal, the $TKE$ is minimal, and, on the contrary, $TKE$ is maximal
where $<R_{12}^{(crit)}>$ is minimal (at $A_H\approx 130$ in the
considered case).
\section{Numerical results}
\label{observa}
Keeping in mind that fission is a {\it slow} process, one could assume
that during the fission process the state of the fissioning
nucleus is close to
statistical  equilibrium, i.e.\
each point $q_i$ on the deformation energy surface is
populated with a probability given by the canonical distribution,
\bel{boltz} P(\alpha_i, q_i)=e^{-\left(\frac{E(\alpha_i, q_i)-Z}{T_{coll}}\right)},Z\equiv -T_{coll}\log\sum_ie^{-\left(\frac{E(\alpha_i, q_i)}{T_{coll}}\right)}.
\end{equation}
Here $T_{coll}$ is a parameter characterizing the width of the distribution \req{boltz} in the space of deformation parameters. The energy
$E(\alpha_i, q_i)$ in \req{boltz} is the sum of the liquid-drop deformation
energy \req{eldm} and of the shell correction $\delta E$, shown in Fig.~\ref{lam56}.
Here, $\alpha_i$ is the mass asymmetry of fissioning system and $q_i$ are the rest of collective parameters (elongation $R_{12}$, and Lagrange multipliers $\lambda_5$ and $\lambda_6$).

The distribution \req{boltz} is a basic assumption of the
scission-point model suggested in \cite{spm} and developed further in \cite{spm2,spm3,spm4},  see also \cite{krapom}.
The parameters of this model were fitted in \cite{spm}
to reproduce the numerous experimental data.
The $T_{coll}$ was found to be close to $1 \mev$.
In the calculations shown below we use somewhat larger
value $T_{coll}$=1.5 MeV.

We are aware that, in principle, the distribution \req{boltz}
should depend not only on the deformation energy but on the
collective velocities too.  The kinetic prescission energy
estimated in this work can reach $20 \mev$ in case of
$^{246}Cm+n_{th}$ reaction. However, even in this case the
collective kinetic energy per particle ($\approx 0.1 \mev$) is
much smaller than the single-particle kinetic energies. So, the
approximation \req{boltz} seems quite reasonable.

The normalized mass distribution of the fission fragments $Y(\alpha)$
can be expressed then in terms of the deformation energy at the critical deformation
$R_{12}^{(crit)}$,
\bel{yield}
Y(\alpha)=\sum_i P(\alpha, R_{12}^{(crit)}, \lambda_{5i}, \lambda_{6i} )\,.
\end{equation}
The summation in \req{yield} is carried out over the Lagrange multipliers $\lambda_{5}, \lambda_{6}$ which correspond to the same $R_{12}^{(crit)}$. The rest of deformation parameters of TCSMP were fixed by the minimization of the liquid-drop energy at $R_{12}=R_{12}^{(crit)}$.
\begin{figure}[ht]
\includegraphics[width=0.9\columnwidth]{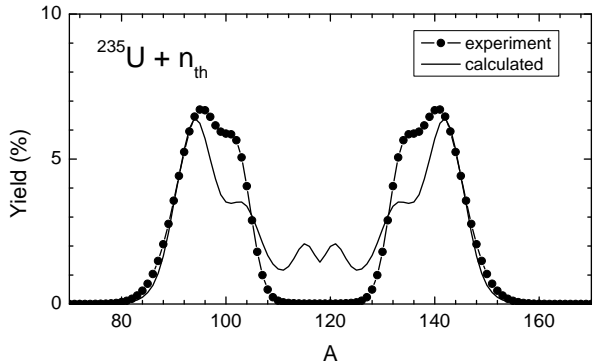}
\caption{The experimental \protect\cite{zeynalov06} and calculated 
values of the mass distribution of fission  fragments in reaction $^{235}U+n_{th}$.} \label{yield236}
\end{figure}

The calculated mass distribution \req{boltz}-\req{yield} of the fission fragments for the
reaction $^{235}U+n_{th}$ is compared with the experimental data
\cite{zeynalov06} in Fig. \ref{yield236}. Please note, that the
information on the deformation energy at the scission point does
not suffice to calculate the width of mass distribution. The
width parameter $T_{coll}$ of the distribution \req{boltz} is a
free parameter in present approach. In Fig. \ref {yield236}, like in all other calculations in this work, we used the value $T_{coll}=1.5 \mev$. The calculated mass distribution is rather close to the experimental results.
The presence and the  position of two mass asymmetric peaks are
reproduced rather well. However, at the symmetric splitting the
calculated values are much larger as compared with experiment.
The reason for such discrepancy is at present not clear.

\begin{figure}[ht]
\includegraphics[width=\columnwidth]{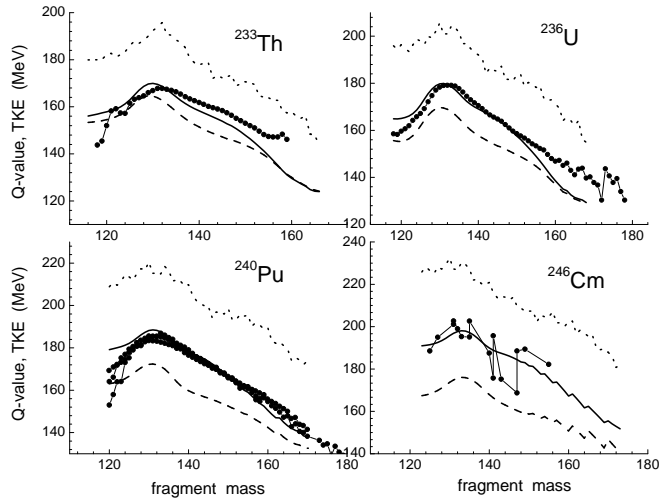}
\caption{The total kinetic energy \protect\req{tke} calculated with (solid) and without (dash) account of fragments prescission kinetic energy. The experiments results (solid circles) are taken from \protect\cite{sergachev,hambsch98,tsuchiya,wagemans,nishio95,rarnas}. The $Q$-values for fission of $^{232}{\rm Th}, ^{235}{\text U}, ^{239}{\rm Pu}$ and $^{245}{\rm Cm}$ are shown by dotted lines.}
\label{qvalue}
\end{figure}

The comparison of the calculated $TKE$ for few fission reaction with the available
experimental results is shown in Fig. \ref{qvalue}. In these
calculations we define $TKE$ as the sum of the Coulomb interaction
of spherical fragments immediately after scission and the
prescission kinetic energy $KE_{pre}$,
\bel{tke}
TKE=<E_{Coul}^{(int)}(\alpha)> + KE_{pre}.
\end{equation}
Here
\bel{Ecoul_mean}
<E_{Coul}^{(int)}(\alpha)>\equiv\frac{1}{2}
\sum_i \frac{Z_LZ_H e^2}{R_{12}^{(crit)}(\alpha,
q_i)}P(\alpha, q_i)/\sum_i P(\alpha, q_i)\,,
\end{equation}
where $eZ_L $ and $eZ_H $ are the charges of light and heavy fragments. The summation in \req{Ecoul_mean} is carried out over $\lambda_{5i}$ and $\lambda_{6i}$.
Like in \cite{fivan14} we define the $KE_{pre}$ from the energy balance
\bel{kepre}
E_{gs}(A_L+A_H)+B_n=E^{(jbs)}+KE_{pre},
\end{equation}
i.e. we assume the "complete acceleration": the energy difference
between the saddle and scission turns into the kinetic energy of
relative motion of fragments, no dissipation takes place. The
opposite extreme case would be the assumption of overdamped
motion: all the energy difference between the saddle and scission
turns into the heat, no prescission kinetic energy.
The incident neutron energy term has been dropped in Eq. \req{kepre} since it very small, $0.025-0.4 ~\text{eV}$.

The
comparison of the calculated and measured values of $TKE$ shown in
Fig. \ref{qvalue} is in favor of "complete acceleration". Without
contribution from $KE_{pre}$ the $TKE$ (dash curve in Fig.
\ref{qvalue}) would be too small.
The neglecting of dissipation for the fission by thermal neutrons
can be also justified by referring to the calculations of the
friction coefficient within the linear response theory. It was
shown in \cite{ivahof} that at small excitations (when the
pairing correlations are still important) the friction
coefficient is negligibly small.

Note, that the calculated $TKE$ is rather close to experimental data.
The position of maximum of $TKE$ and the drop at symmetric
splitting are also well reproduced. This can be considered as a
confirmation that the potential energy just before scission and
the mean value of $<R_{12}^{(crit)}>$ are defined correctly.

The $TKE$ is one of two parts of the total energy release $Q$,
$TKE+TXE=Q$. The $Q$-value is the difference of the ground state
energies \bel{qval} Q\equiv
E_{gs}(A_L+A_H)+B_n-E_{gs}(A_L)-E_{gs}(A_H)
\end{equation}
To calculate the ground state energy $E_{gs}$ besides mass numbers of fission fragments $A_L$ (or $A_H$) one should know also their charge numbers $Z_L$ (or $Z_H$). In the calculations shown in Fig. \ref{qvalue} we considered a Gaussian  distribution $P(Z)$ of charge numbers \cite{wahl} characterized by the most probable charge $Z_p$ and the width parameter $c$,
\bel{charge}
P(Z)=\frac{1}{\sqrt{c\pi}}e^{-(Z-Z_p)^2/c}  .
\end{equation}
The value of parameters $Z_p$ and $c$ was taken from
\cite{litaize}, $c=2(\sigma_Z^2+1/12)$, with $\sigma_Z=0.59$ and,
for heavy fragment,  $Z_p=Int(Z A_H/A)$ with $A$ and $Z$ being
the mass and charge number of mother nucleus. Here $Int(x)$
stands for the integer part of $x$.

\begin{figure}[ht]
\includegraphics[width=0.8\columnwidth]{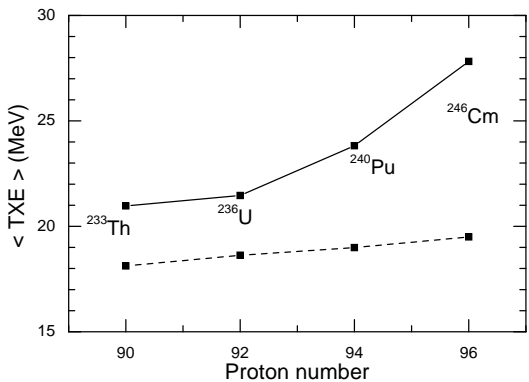}
\caption{The total excitation energy
\protect\req{txeavr} (solid) and the jump of liquid-drop energy
during the neck rupture \protect\req{deledef}(dash).}\label{txe}
\end{figure}

The $Q$-value does not depend on dynamics, it can be calculated
within the macroscopic-microscopic method or taken from the
existing databases. Since the experimental value of $TKE$ is rather
well reproduced by the present calculations, the calculated total
excitation energy $TXE=Q-TKE$ should be also quite accurate. The average
value of TXE
\bel{txeavr}
<TXE>=\sum_{A_H}[Q(A_H)-TKE(A_H)]Y(A_H)
\end{equation}
is shown in Fig. \ref{txe}.
For the comparison the excitation energy due to the difference $\Delta E_{def}^{LD}$ of the liquid-drop energy \req{deledef} just before the scission and immediately after the scission is plotted by dash line. The contribution of $\Delta E_{def}^{LD}$ to the total excitation energy of fission fragments varies from 85 $\%$  (for $^{233}{\rm Th}$) to 60 $\%$  (for $^{246}{\rm Cm}$). The rest is the contribution from the shell effects.

The
fragments are de-excited by the emission of neutrons and
$\gamma$-rays. When the excitation energy is smaller than neutron
separation energy $S_n$ the $\gamma$-quanta are emitted. The
energy available for $\gamma$-emission varies from $S_n$ to zero.
So, on average the excitation energy available for neutron
emission is given by $E_x-S_n/2$. The average value of the total
number $\bar{\nu}_{tot}$ of prompt neutrons can be estimated by the relation
\bel{nuavr}
\bar{\nu}_{tot}\approx <TXE> / \bar S_n -1/2,\, \text{with}\,\, \bar S_n=5.7 \mev .
\end{equation}
In order to calculate the dependence of $\bar{\nu}$ on the fragments mass number $A$
 one needs to know how the excitation
energy is shared between the fragments. Here we split the last
stage of the fission into two steps. On first step the rapid neck
rupture takes place.
The nucleus turns into two fragments with
the same mass asymmetry and the same distance between centers of
mass as just before scission. The shape of the fragments
immediately after scission is assumed to correspond to the
minimum of the liquid drop energy. It was shown in \cite{procedia} that
the optimal shape of the fragments placed at the distance
$R_{12}^{(crit)}$ is very close to two spheres. The contribution
to energy from quadrupole and higher order deformation is smaller
that $0.5 \mev$ and can be neglected. So, in present work we
assume that immediately after scission the two spherical
fragments are formed. The energy immediately after scission
consists of the energy of light and heavy spherical
fragments plus the energy of their Coulomb interaction,
\bel{eias}
E^{(ias)}(A)=E^{(sph)}(A_L)+E^{(sph)}(A_H)+E_{Coul}^{(int)}(R_{12},
\alpha).
\end{equation}

\begin{figure}[ht]
\includegraphics[width=\columnwidth]{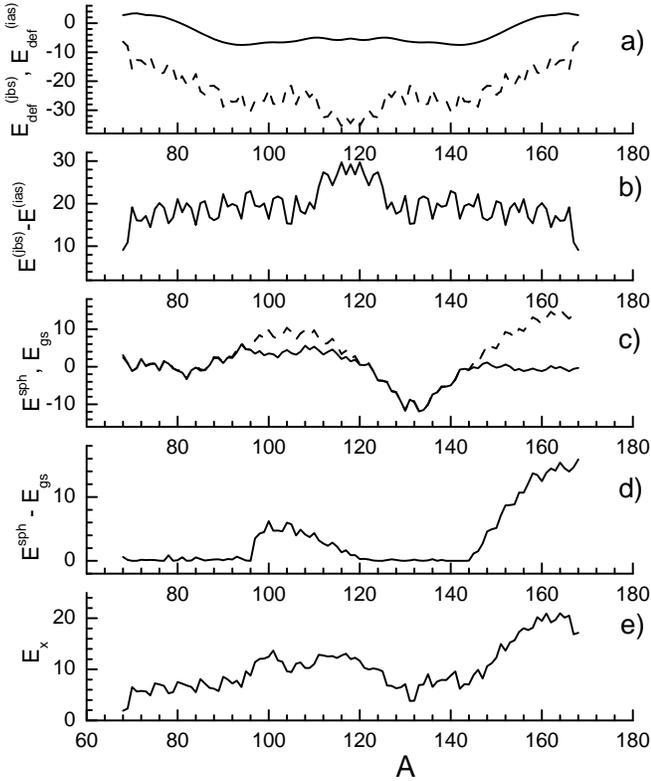}
\caption{ a) the deformation energy of $^{236}{\text U}$ just before scission (solid) and immediately after scission (dash) as
function of fragments mass number $A$; b) the primary excitation
energy $E_{def}^{jbs}-E_{def}^{ias}$; c) the deformation energy
at spherical shape (dash) and at the ground state (solid); d)
the extra deformation energy \protect\req{deledef} of fission
fragments; e) the excitation energy \protect\req{exx} of fission
fragment. All energies in this figure are given in MeV.}
\label{five}
\end{figure}

The primary excitation energy $E^{(ias)}-E^{(jbs)}$ is shared between the
fragments. In the recent work \cite{cahaivta} the partition of the
excitation energy between the light and the heavy fragments
$E^*_{ias}(L(H))$ was calculated both in the thermalization
immediately after scission hypothesis ($T_L=T_H$) \cite{madnix}
and in the sudden approximation \cite{caharise}. The  effect of
these two hypothesis on the  mass split is quite opposite. While
the sudden approximation predicts the lighter fragment to be more
excited, the contrary is  in the case of thermalization. The
difference of the excitation energy of light and heavy fragment
is not so large, only few MeV. So, for simplicity we will assume
here that the excitation energy of light and heavy fragments due
to the neck rupture is the same. This excitation energy for
$^{236}{\text U}$ is shown in Fig. \ref{five}b.

On the second stage of fission process the fragments relax to the
ground state shape and gain some extra deformation energy
\bel{deledef1} \Delta E_{def}(A)=E^{(sph)}(A)-E_{gs}(A)
\end{equation}
Extra deformation energy $\Delta E_{def}$ is shown in Fig. \ref{five}d
as function of fragments mass number. Eventually, the total excitation energy of one fragment is given by the sum
\bel{exx}
E_x(A)=[E^{(jbs)}(A)-E^{(ias)}(A)]/2+\Delta E_{def}(A).
\end{equation}
see Fig. \ref{five}e.

\begin{figure}[ht]
\includegraphics[width=\columnwidth]{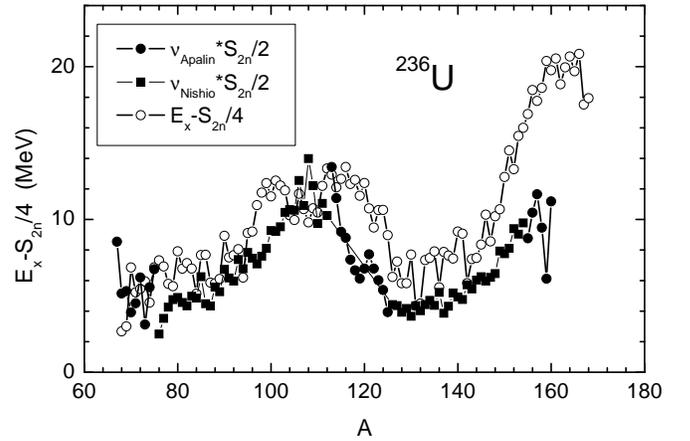}
\caption{The excitation energy \protect\req{exx} available for
prompt neutron emission (open circles) and the experimental
results for neutron multiplicity \protect\cite{apalin,nishio}
multiplied by half of two-neutron separation energy.} \label{sawtooth}
\end{figure}

In Fig. \ref{sawtooth} we compare the excitation energy available
for the prompt neutron emission, $E_x-0.5 S_n$ with the
experimental value of neutron multiplicity \cite{apalin,nishio}
multiplied by the half of two-neutron separation energy $S_{2n}$ (in order to remove the rapid fluctuations due to the odd-even effect in $S_n$). One can see
that there is some discrepancy up to 5 $\mev$ at large mass
asymmetries, but on the average the saw-tooth structure is rather
well reproduced.

Having at one's disposal the excitation energy $E_x(A)$ one can calculate the total excitation energy
\bel{exxtot}
<E_x>=\sum_{A}E_x(A)Y(A).
\end{equation}
and the total number of prompt neutrons
\bel{nu}
\bar{\nu}_{tot}=\sum_{A}[E_x(A)/S_n(A)-1/2]Y(A).
\end{equation}
The simple approximation to \req{nu} is given by \req{nuavr}. In \req{nu} the multiplicity of emitted neutrons is calculated for each value of $A_H$ and summed in $A_H$ with the weight given by the calculated mass distribution of fission fragments. In \req{nuavr} the total neutron multiplicity is defined as the ratio of the total excitation energy to the average neutron separation energy.

\begin{figure}[hb]
\includegraphics[width=0.8\columnwidth]{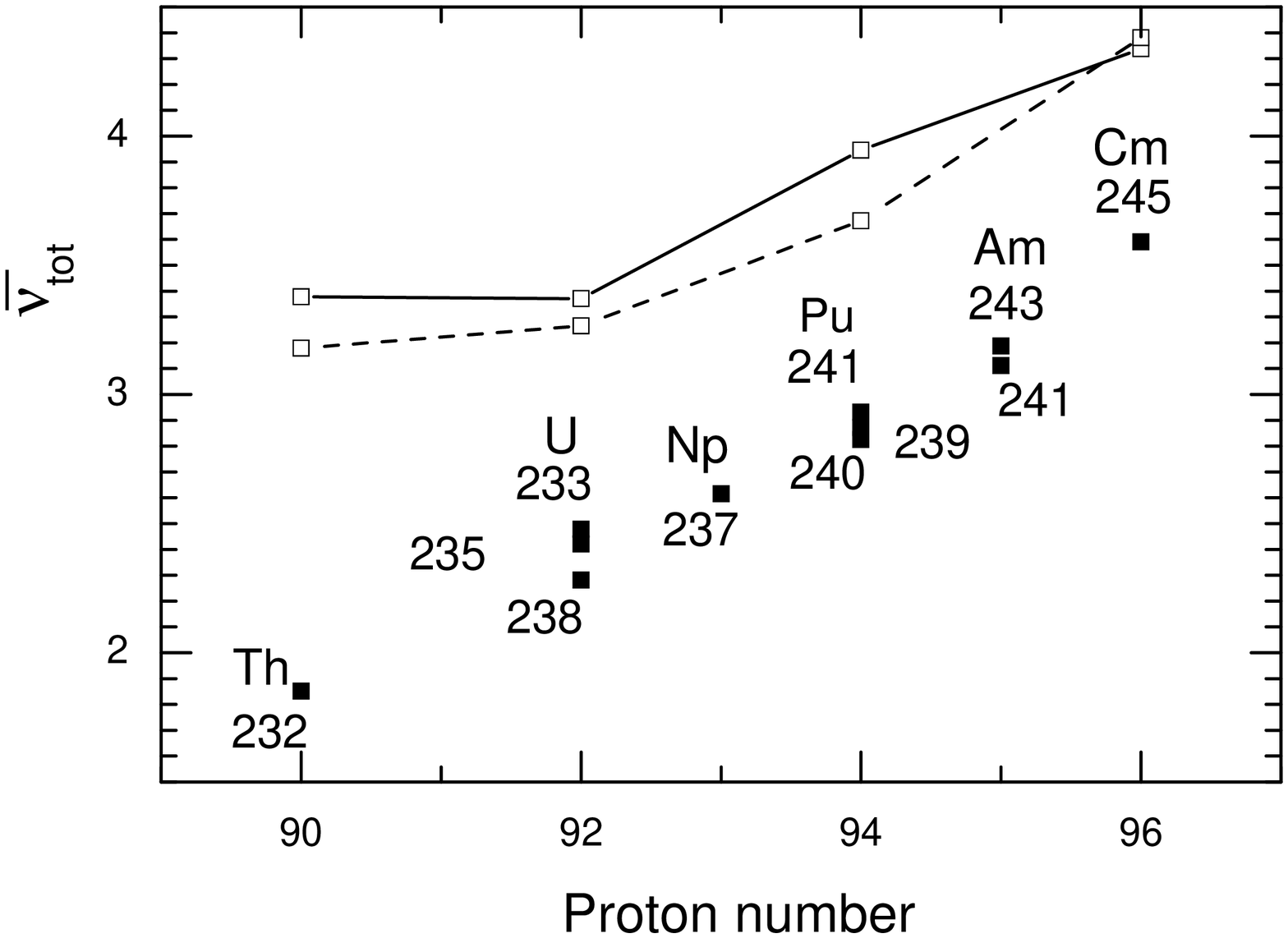}
\caption{The calculated value of total neutron multiplicity
\protect\req{nu} (open squares) and the experimental results
\protect\cite{ohsawa} (solid squares). The dash line shows the
estimate \protect\req{nuavr}.}\label{nutotc}
\end{figure}

The comparison of calculated $\bar{\nu}_{tot}$  with the available
experimental results is shown in  Fig. \ref{nutotc}.
On average
the dependence of $\bar{\nu}_{tot}$ on the proton number of fissioning nuclei
is qualitatively reproduced. The calculated value of $\bar{\nu}_{tot}$
are however larger than experimental by (0.5-0.9). The source of
this discrepancy can be related to the use of very simple estimate \req{nuavr} for the neutron multiplicity and use of the quasistatic
approximation \req{boltz} for the mass distribution. In
particular, too large value of $Y(A)$ for symmetric splitting, see Fig. \ref{five}e is
not confirmed by the experimental results.

\section{Summary}
\label{summa}
The calculations carried out in present work show that the
optimal shape prescription offers a good possibility to define
the shape of fissioning nuclei just before the scission. This
information can be used for the evaluation of the quantities
which are measured in fission experiments like the mass distribution, the total kinetic and
excitation energy of fission fragments, the multiplicity of prompt neutrons. The calculated
distributions of the total kinetic energy for fission of
$^{232}{\rm Th}, ^{235}{\text U}, ^{239}{\rm Pu}$ and $^{245}{\rm Cm}$ were found to be in rather good
agreement with the experimental data.

The sawtooth structure of the neutron multiplicity is well reproduced.
 The dependence of total
multiplicity of prompt neutrons on proton number of fissioning
nucleus is reproduced only qualitatively.

The reason for the discrepancy between calculated and experimental results
for average total prompt neutron multiplicities may be too crude approximation of
Eqs. 26, 31 for the relation between the excitation energy and neutron multiplicity.
There could be also other reasons, like the use of assumption  that each point on the deformation energy surface is populated with a probability given by the canonical distribution \req{boltz}.

For the mass distribution of fission fragments
the position of the peaks is reproduced rather well. The width
and strength of peaks of calculated mass distribution differ
substantially from the experimental data. For more accurate description of mass distribution the dynamical approach to the fission process seems necessary.

\begin{acknowledgments}
This paper includes the results of ``Comprehensive study of delayed-neutron yields for accurate
evaluation of kinetics of high-burn up reactors"
entrusted to Tokyo Institute of Technology by the Ministry of Education, Culture, Sports, Science 
and Technology of Japan (MEXT).
The authors appreciate very much the fruitful discussions with Profs. N. Carjan, A. Iwamoto, K. Nishio and K. Pomorski.
One of us (F. I.) would like to express his gratitude to the
Tokyo Institute of Technology, for the 
hospitality during his stay at Japan.
\end{acknowledgments}


\end{document}